\documentclass{osa-article}

\journal{oe}

\articletype{Research Article}

\begin{document}

\title{Ultra-wideband THz/IR Metamaterial Absorber based on Doped Silicon}

\author{Huafeng Liu,\authormark{1,2} Kai Luo,\authormark{3} Danhua Peng,\authormark{1,2} Fangjing Hu,\authormark{1,2,\authormark{*}} and Liangcheng Tu\authormark{1,2}}

\address{\authormark{1}MOE Key Laboratory of Fundamental Physical Quantities Measurement, Huazhong University of Science and Technology, Wuhan, 430074, People's Republic of China\\
\authormark{2}Hubei Key Laboratory of Gravitation and Quantum Physics, School of Physics, Huazhong University of Science and Technology, Wuhan, 430074, People's Republic of China\\
\authormark{3}School of Electronic Information and Communications, Huazhong University of Science and Technology, Wuhan, 430074, People's Republic of China}
\email{\authormark{*}fangjing{\textunderscore}hu@hust.edu.cn}



\begin{abstract}
Metamaterial-based absorbers have been extensively investigated in the terahertz (THz) range with ever increasing performances. In this paper, we propose an all-dielectric THz absorber based on doped silicon. The unit cell consists of a silicon cross resonator with an internal cross-shaped air cavity. Numerical results suggest that the proposed absorber can operate from THz to mid-infrared, having an average power absorption of $\sim$95{\%} between 0.6 and 10 THz. Experimental results using THz time-domain spectroscopy show a good agreement with simulations. The underlying mechanisms for broadband absorptions are attributed to the combined effects of multiple cavities modes formed by silicon resonators and bulk absorption in the substrate, as confirmed by simulated field patterns. This ultra-wideband absorption is polarization insensitive and can operate across a wide range of the incident angle. The proposed absorber can be readily integrated into silicon-based platforms and is expected to be used in sensing, imaging, energy harvesting and wireless communications systems. 
\end{abstract}

\section{Introduction}
Terahertz (THz) absorbers with broadband operations are essential components for various applications such as sensing, imaging, energy harvesting and wireless communications. Although some materials in nature have shown reasonably good absorptions within the THz regime, it is still in great demand of ultra-wideband absorbers with flat and high power absorptance.  Metamaterial-based perfect absorbers (MPAs) have been extensively studied since it was first proposed in the microwave range by Landy \textit{et al} \cite{PhysRevLett.100.207402}. Over the past years, the performances of MPAs have been increased and the operating frequencies have been expanded from microwaves to THz, infrared and visible ranges due to excellent scalability of metamaterials \cite{doi:10.1002/adma.201200674,Tao:08,PhysRevB.79.045131,doi:10.1063/1.3442904,Aydin:2011}. 

The unit cell of conventional MPAs usually consists of a metallic resonator, a dielectric spacer and a ground layer. At its resonant frequency, near-unity power absorptance can be achieved as both the power transmittance and reflectance are minimised. However, MPAs with metallic resonators normally exhibit narrow bandwidths because of their resonating nature. For broadband operation, either complicated unit cells shapes, or composite and multilayer structures are required, limiting the applications for the increased fabrication and design complexities. To tackle this problem, doped silicon substrates have been used to create broadband MPAs. Different unit cells patterns, including circular holes \cite{Kakimi:2014}, rectangular cubes \cite{Pu:12,doi:10.1063/1.4890617,doi:10.1063/1.4929151}, sawtooth structures \cite{doi:10.1063/1.4950800}, crosses \cite{7331678, doi:10.1002/adom.201400368} and dumbbell shapes \cite{Zang:2015}, have been demonstrated. The broadband absorptions are mainly due to the creation of plasmonic waveguide modes or by multi-interference and diffraction effects. The electromagnetic (EM) responses of the devices can be further engineered by changing the parameters of the unit cell or by adjusting the doping concentration. 

Inspired by these previous studies, in this paper, we propose an ultra-wideband absorber based on a standard 400 $\mu$m thick doped silicon substrate and investigate its electromagnetic responses in the THz and mid-infrared spectra. The unit cell consists of a silicon cross structure with an internal cross-shaped air cavity, creating multiple air-cavity modes to reduce the reflection over a large bandwidth. Numerical results suggest an average power absorptance of $\sim$95{\%} from 0.6 to 10 THz, and a consistent performance across a wide range of the incident angle for both TE and TM polarisations can be obtained. Terahertz time-domain spectroscopy (THz-TDS) measurements from 0.2 to 2.6 THz show a good agreement with simulated results. 

\section{Principle and design}

A 400 $\mu$m thick doped silicon substrate with a resistivity of 0.02 $\Omega$$\cdot$cm is used to fabricate the absorber. The complex dielectric constant of silicon can be described by Drude model as 

\[\epsilon = \epsilon_{\infty} - \frac{{\omega_p}^2}{{\omega}^2 + j \omega \gamma}\]
where $\epsilon_{\infty}$ = 11.68 is the permittivity of silicon at high frequencies, $\omega_p$ = 2$\pi$$\times$7.88 THz is the plasma frequency, and $\gamma = 2\pi \times 1.78$ THz is the collision frequency \cite{doi:10.1002/adom.201400368}.

\begin{figure}[ht]
\centering\includegraphics[width=7cm]{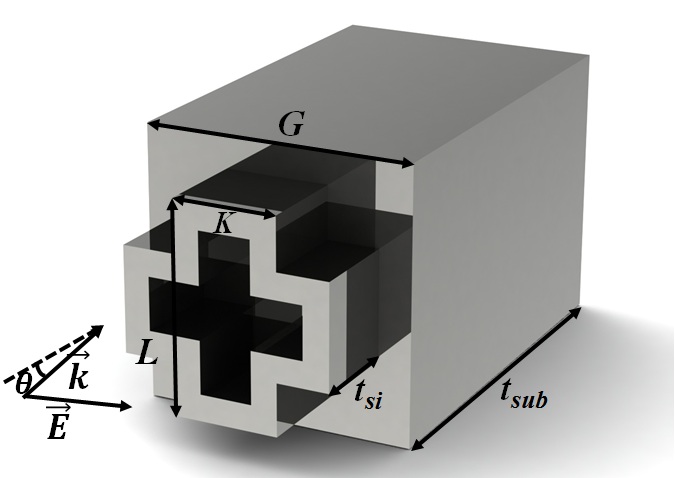}
\caption{Schematic drawing of the unit cell for the proposed THz absorber.}
\label{fig:schematic}
\end{figure}

As shown in Figure \ref{fig:schematic}, the unit cell of the proposed absorber consists of a silicon cross structure with an internal cross-shaped air cavity. To obtain the EM responses and gain an insight into the underlying mechanism for broadband absorption, full-wave simulations were performed using the unit cell boundary condition in CST Microwave Studio. The incident wave is TE-polarised and the propagation direction is also shown in Figure \ref{fig:schematic}, where $\theta$ is the incident angle. Subsequently, the power transmittance $T(\omega)=|S_{21}(\omega)|^2$, reflectance $R(\omega)=|S_{11}(\omega)|^2$ and absorptance $A(\omega)=1-T(\omega)-R(\omega)$ can be obtained.

\begin{figure}[ht]
\centering\includegraphics[width=13cm]{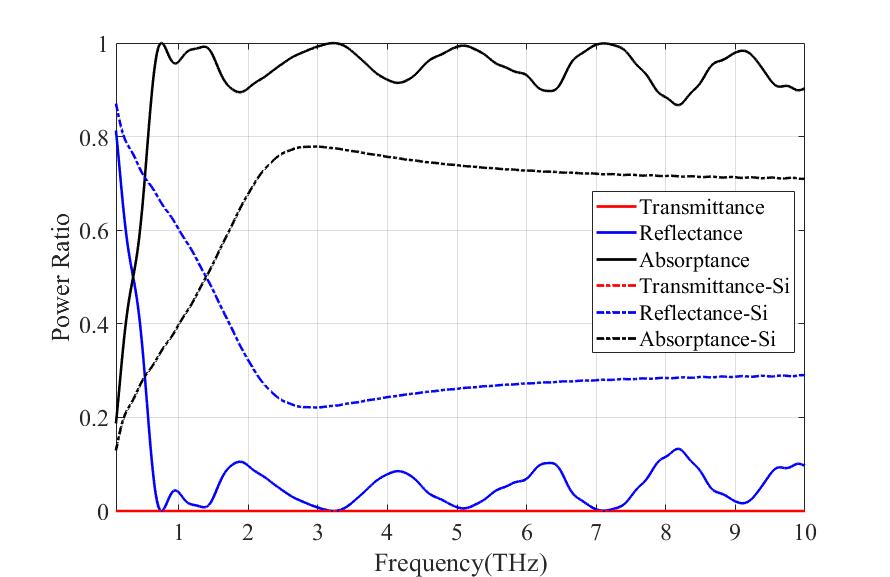}
\caption{Simulated results for the proposed absorber and a bare silicon substrate.}
\label{fig:simulated_results}
\end{figure}

Figure \ref{fig:simulated_results} shows the power responses of the proposed THz absorber, as well as a unpatterned silicon substrate of the same 400 $\mu$m thickness. Here, optimized values are $G$ = 210 $\mu$m, $L$ = 160 $\mu$m, $K$ = 80 $\mu$m, $t_{Si}$ = 75 $\mu$m and $t_{Sub}$ = 325 $\mu$m. It is seen that the power absorptance for an unpatterned silicon wafer improves as the frequency increases, reaching to its maximum of $\sim$78$\%$ at 2.8 THz, and then starts decreasing to about 70\%. For the proposed absorber, the power absorptance becomes greater than 90\% from 0.6 THz, and sustains a high value across the entire spectrum of interest up to 10 THz. The calculated average absorptance within 0.6-10 THz is $\sim$95\%. Within this bandwidth, the transmitted power can be neglected for $t_{Sub}$ = 325 $\mu$m.

\begin{figure}[ht]
\centering\includegraphics[width=13.2cm]{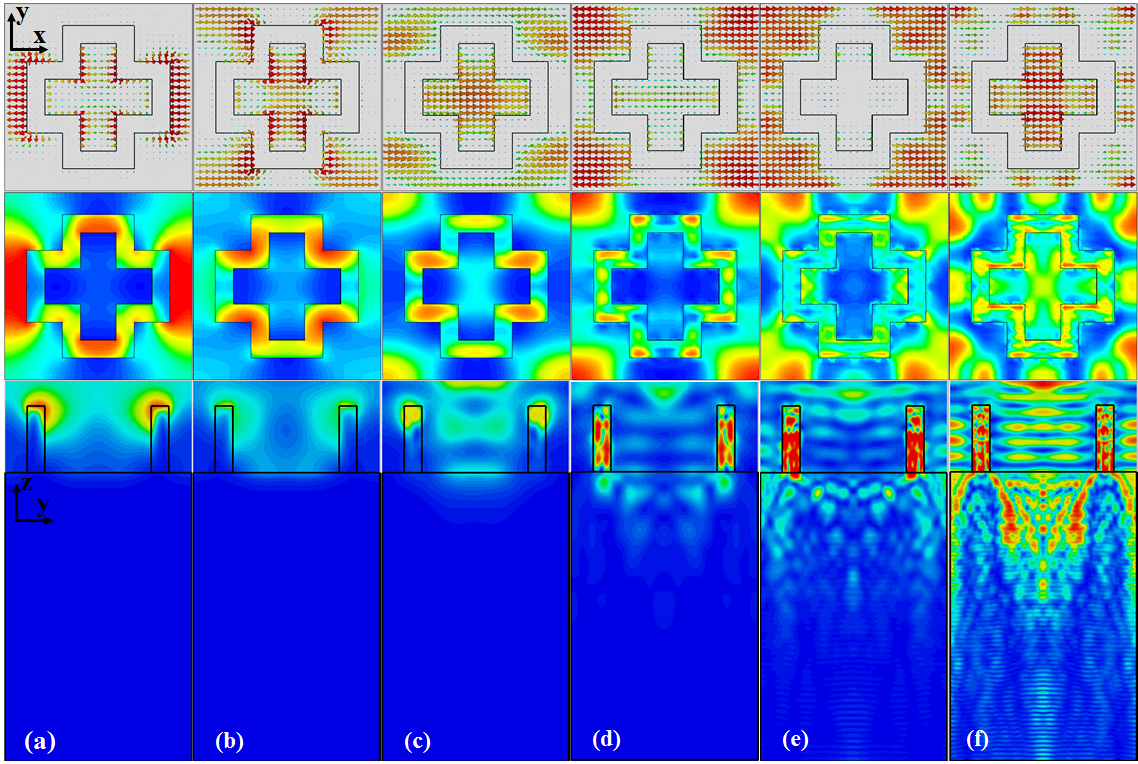}
\caption{Simulated electric and magnetic field patterns for the absorber at (a) 0.75 THz, (b) 1.38 THz, (c) 3.23 THz, (d) 5.10 THz, (e) 7.13 THz and (f) 9.12 THz, respectively.}
\label{fig:magnetic_field}
\end{figure}

In order to investigate the origins of these absorption peaks, instantaneous electric fields and average magnetic fields at the resonant frequencies are shown in Figure \ref{fig:magnetic_field}. The first two rows illustrate the top view and the last row is the side view of the $yoz$ plane across the centre of the unit cell. It is clearly demonstrated that distinct field patterns were obtained at different resonant frequencies. At low frequencies, these absorption peaks were originated by different cavity modes within the unit cell. For example, the electric fields at the first resonance of 0.75 THz were localized between the left and right arms of the adjacent crosses, which can be treated as a parallel-plate plasmonic waveguide \cite{Maier:2007:10.1007/0-387-37825-1}. Figure \ref{fig:magnetic_field} (b) shows that the most of the energy was absorbed within the four corners of the cross resonators. As the frequency increases, more cavity modes were observed, and field patterns became more complicated.  At 5.10 and 7.13 THz, the TE-like modes within the rectangular cavities formed by the four corners of the adjacent cross resonators became dominant. At the highest evaluated resonant frequency of 9.12 THz, more incident energy was able to propagate through the top structure and absorbed by the silicon substrate, as evidenced in Figure \ref{fig:simulated_results}(f). This shows the effectiveness of applying thick doped silicon substrate to achieve a wider operating bandwidth.

\section{Fabrication and Experiment}

\begin{figure}[ht]
\centering\includegraphics[width=8cm]{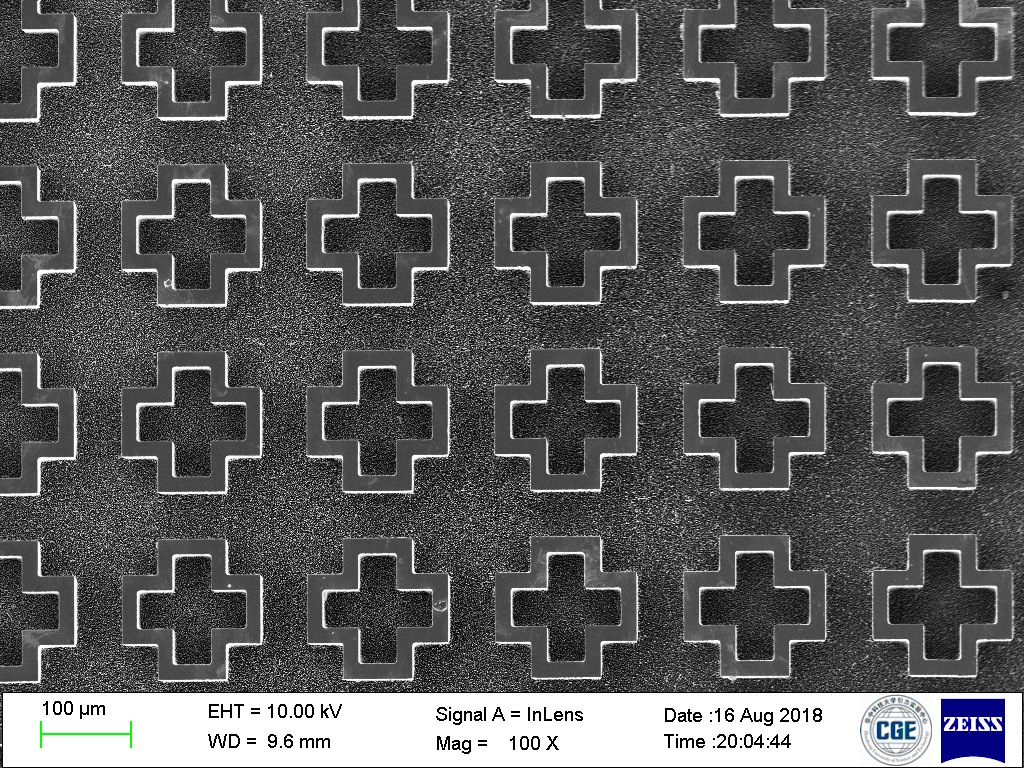}
\caption{SEM image of the fabricated THz absorber.}
\label{fig:SEM}
\end{figure}

The proposed design was fabricated using a 4-inch 400 $\mu$m thick silicon wafer. First, a 5 $\mu$m thick AZ9260 photoresist layer was spin-coated on the primer prepared wafer surface. Then, the wafer was exposed by ultraviolet (UV) light through the designed photomask. After the development process in the developer AZ400K, the exposed areas of the photoresist layer were striped, followed by a standard deep reactive ion etching (DRIE) process to etch silicon through the photoresist-free windows. Once the DRIE process is finished, the wafer was socked into acetone solution for at least 30 minutes to strip organic outgrowths and the residual photoresist. Finally, the 75 $\mu$m deep silicon trenches were formed on the substrate. The overall size of the fabricated sample chip was 19.95 $\times$ 19.95 mm$^2$ with 95$\times$95 (9025) unit cells. A scanning electron microscope (SEM) image of the sample is shown in Figure \ref{fig:SEM}. 

Both transmission and reflection measurements were performed at normal incidence (TE-polarisation) using a Zomega Z-3 THz time-domain spectrometer with a spectral coverage from 0.2 to 2.6 THz. Measured results agree well with simulated values, having the designed resonant frequencies clearly identified. There is a slight redshift at the first resonance of 0.75 THz, and a reduced power absorptance near 1 THz. The discrepancies are mainly due to fabrication errors of the sample. Higher absorptions can be achieved by employing additional capping layers \cite{doi:10.1063/1.4929151} to match the impedance of air and silicon layers, and can be scaled to other frequencies by changing the doping concentration \cite{Pu:12}.

\begin{figure}[ht]
\centering\includegraphics[width=13cm]{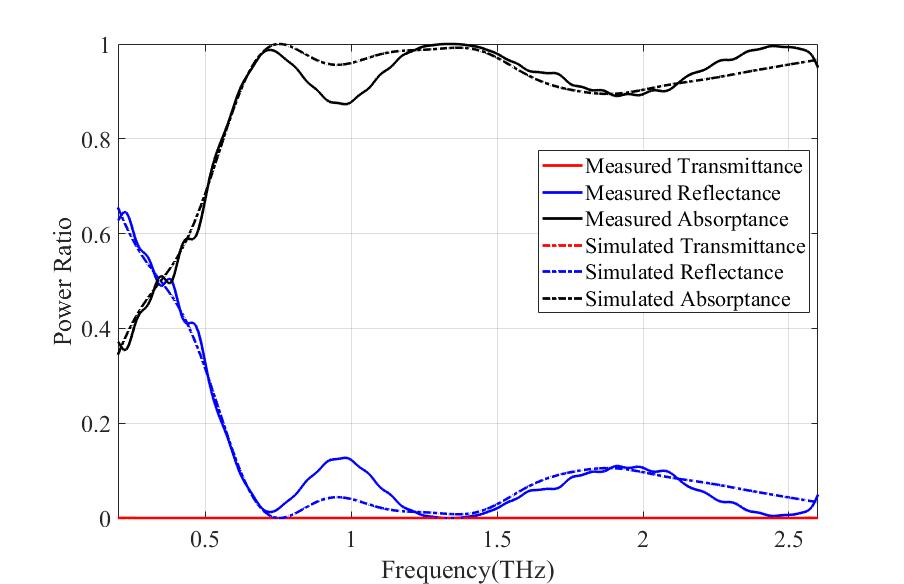}
\caption{Measured power responses for the proposed THz absorber.}
\label{fig:measured_results}
\end{figure}

\begin{figure}[ht]
\centering\includegraphics[width=13cm]{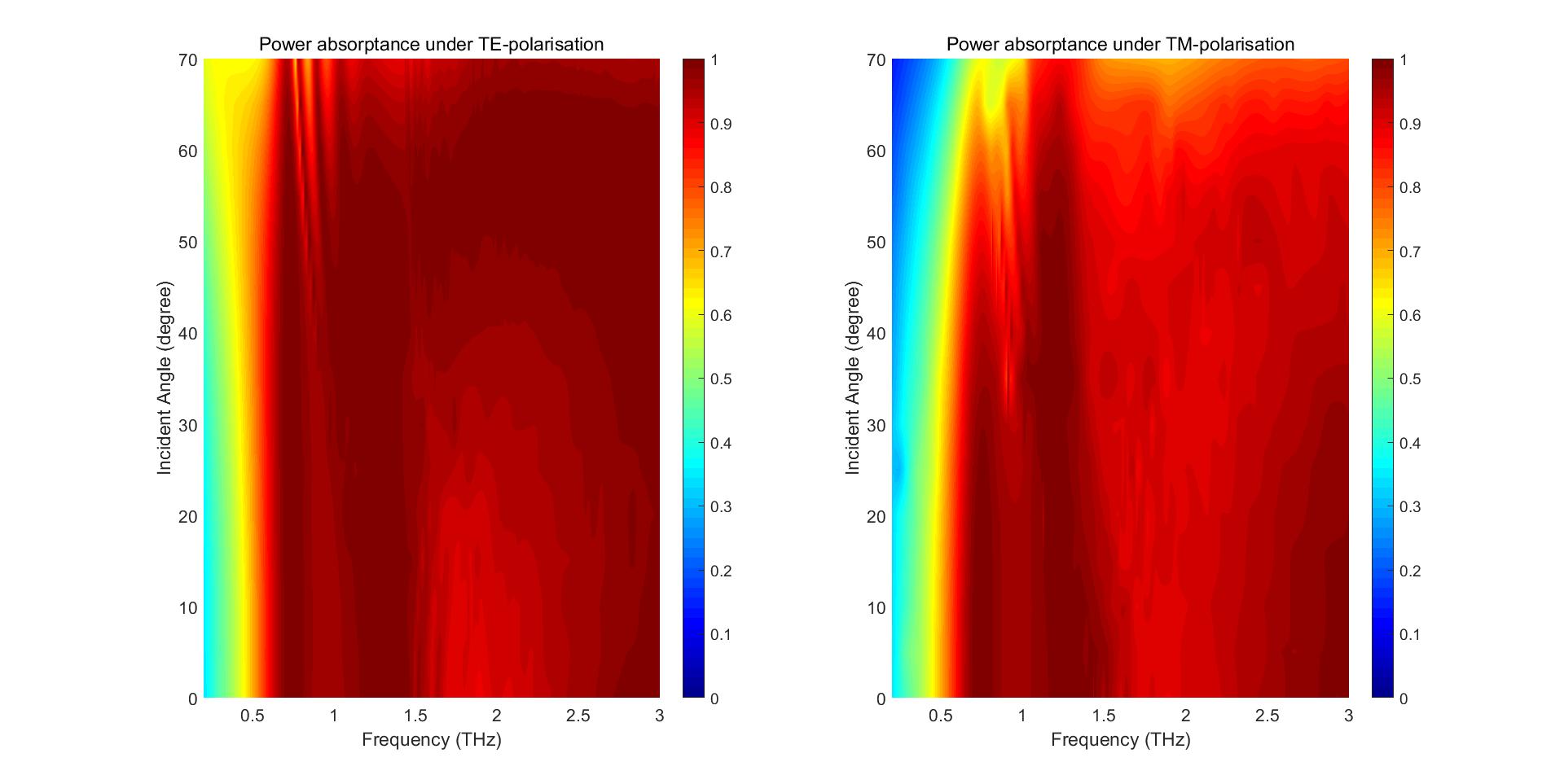}
\caption{Simulated power absorptance as a function of frequency and incident angle.}
\label{fig:angle}
\end{figure}

As a four-fold symmetric structure, the proposed absorber is polarisation insensitive at normal incidence under both TE- and TM-polarisazitons. At oblique incidence, this structure is dependent on the polrisation as well as the incident angle due to the power imbalance within the cavity structures. Figure \ref{fig:angle} shows the simulated power absorptance under different polarisations for the proposed THz absorber, as a function of frequency (0.2-3 THz) and the incident angle ($0^\circ$-$70^\circ$). For TE-polarisation, the power absorption remains greater than 90\% from 1 to 3 THz for incident angles up to $70^\circ$.  Between 0.6 and 1 THz, the absorptance decreases to about 70\% as the incident angle increases. The dependencies of the power absorptance spectra on the incident angle become more significant for TM-polarisation. From 1 to 3 THz, the absorptance remains greater than 90\% for incident angles up to $50^\circ$ and then decreases to about 80\% for a $65^\circ$ incidence.

\section{Conclusion}
In conclusion, we have designed, fabricated and experimentally demonstrated an ultra-wideband absorber within the THz and mid-infrared spectra. The absorber was fabricated based on a standard 400 $\mu$m thick doped silicon substrate. Each unit cell contains a 75 $\mu$m thick silicon cross structure with an inner cross-shaped air cavity. The ultra-wideband absorption was achieved due to the combined effects of multiple cavity modes and bulk absorption of the silicon substrate. At low frequencies, different cavity modes were formed within this structure, significantly increasing the absorption when compared to unpatterned silicon substrate of the same thickness. At high frequencies, more energy was absorber by the silicon substrate itself. The average power absorptance was simulated to be $\sim$95\% from 0.6 to 10 THz and agrees well with the THz-TDS measurements between 0.2 and 2.6 THz. This absorber is polarisation insensitive, and can sustain a high power absorptance for incident angles up to 50$^{\circ}$ under both polarisations. More importantly, it can be readily integrated into other silicon-based platforms, and can be used in sensing, imaging, wireless communications, thermal emitting and energy harvesting systems.

\section*{Acknowledgments}
This work was partially supported by the National Key R\&D Program of China (Grant No. 2018YFC0603301), the Natural Science Foundation of China (Grant No. 61801185), and HUST Key Innovation Team Foundation for Interdisciplinary Promotion (Grant No. 2016JCTD102).


\bibliography{sample}

\end{document}